\documentclass[a4paper,11pt,notitlepage]{article}

\newcommand*\diff{\mathop{}\!\mathrm{d}}

\usepackage[margin=1in]{geometry}
\usepackage{apacite}
\usepackage{natbib}
\usepackage{csquotes}
\usepackage{import}
\usepackage{soul}
\usepackage{amssymb}
\usepackage{amsmath}
\usepackage{pifont}
\usepackage{xcolor}
\usepackage{bbm}
\usepackage{bm}
\usepackage{dsfont}
\usepackage[ruled,vlined]{algorithm2e}

\usepackage{authblk}
\usepackage{hyperref}
\usepackage{graphicx}
\usepackage[title,titletoc]{appendix}

\usepackage{wrapfig}
\usepackage{titling}
\usepackage{float}

\newtheorem{prop}{Proposition}

\newcommand\independent{\protect\mathpalette{\protect\independenT}{\perp}}
\def\independenT#1#2{\mathrel{\rlap{$#1#2$}\mkern2mu{#1#2}}}

\usepackage{xr}
\title{Modelling intransitivity in pairwise comparisons with application to baseball data} 
\author{Jess  Spearing$^1$, Jonathan Tawn$^1$, David Irons$^2$, Tim Paulden$^2$\\
{\small $^1$ Lancaster University, $^2$ ATASS Sports} }
\date{\today}
\begin{document}
\maketitle
\begin{abstract}
The seminal Bradley-Terry model exhibits transitivity, i.e., the property that the probabilities of player A beating B and B beating C give the probability of A beating C, with these probabilities determined by a skill parameter for each player. Such transitive models do not account for different strategies of play between each pair of players, which gives rise to {\it intransitivity}. Various intransitive parametric models have been proposed but they lack the flexibility to cover the different strategies across $n$ players, with the $O(n^2)$  values of intransitivity  modelled using $O(n)$ parameters, whilst they are not parsimonious when the intransitivity is simple. We overcome their lack of adaptability by allocating each pair of players to one of a random number of $K$ intransitivity levels, each level representing a different strategy.  Our novel approach for the skill parameters involves having the $n$ players allocated to a random number of $A<n$ distinct skill levels, to improve efficiency and avoid false rankings.
Although we may have to estimate up to $O(n^2)$ unknown parameters for $(A,K)$ we anticipate that in many practical contexts $A+K < n$. Our semi-parametric model, which gives the  Bradley-Terry model when $(A=n-1, K=0)$, is shown to have an improved fit relative to  the Bradley-Terry, and the existing intransitivity models, in out-of-sample testing when applied to simulated and American League baseball data. Supplementary materials for the article are available online.
\end{abstract}

\section{Introduction}
\label{sec:intro}
%{\color{red}
The seminal Bradley-Terry model \citep{bradley1952rank} is commonly used to rank objects from paired comparison data. Given a set $\mathcal{I}$ of $n$ objects with each object $i \in \mathcal{I}$ having skill $r_i \in \mathbb{R}$, then the Bradley-Terry model gives, for $i\neq j\in \mathcal{I}$, 
\begin{equation}
\label{eq:BT_definition}
p^{(\text{BT})}_{ij} = {\Pr\{i\succ j\} := \{1 + \exp[-(r_i-r_j)]\}^{-1}},
\end{equation}
where $a \succ b$ denotes preference for object $a$ over $b$, and $r_1=0$ to avoid identifiability issues. A ranking of the objects is given by sorting estimates of $r := \{r_i \in \mathbb{R} : i \in \mathcal{I}\}$. This model is transitive, i.e., 
$p^{(\text{BT})}_{jk}$ is given by 
$p^{(\text{BT})}_{ij}$ and $p^{(\text{BT})}_{ik}$, 
for all  $i\neq j \neq k \in \mathcal{I}$, see Section~\ref{sec:model}.

Now consider the game of Rock-Paper-Scissors, a zero-sum game in which Rock beats Scissors, Scissors beats Paper, and Paper beats Rock, and specifically consider the deterministic scenario where players (r,p,s) always pick (Rock, Paper, Scissors) respectively.
In this scenario, all win probabilities in a game are either $0$ or $1$ depending on the opponent, and each player wins their next game with probability $1/2$ if their next opponent is to be selected at random. Whatever way the skill of a player is defined, the symmetry of this game set-up unquestionably leads to the conclusion that the three players have equal skill levels.

Conclusions drawn from a Bradley-Terry model fitted to data from this simple game are surprisingly poor.  Given a round-robin tournament, where each player plays all other players an equal number of times, the model will correctly estimate that all players are equally ranked in terms of skills; however, it would also estimate all pairwise win probabilities to be 1/2, which couldn't be more wrong. Even worse, is that any illusory ranking can result when the tournament is not round-robin, e.g., if the most common pairing of players is (r,s) and the other two pairings occur equally often then the Bradley-Terry model will rank player r as top. The key reason for the failure of the Bradley-Terry model is its transitive nature, a trait shared by almost all commonly used ranking systems. 

Here we develop a novel pairwise comparison model, and an associated ranking system, which accounts for {\it intransitivity}. Thus, it describes how specific pairwise probabilities differ from probabilities given by overall skill levels alone, i.e., how probabilities differ from those given by the Bradley-Terry model. The Rock-Paper-Scissors game also illustrates that ranking can involve ties, where subsets of players can have equal skill levels, and that
tournament structure can effect the subsequent inference. We also address some aspects associated with these issues.

The concept and associated modelling of intransitivity is not new. \cite{makowski2006quantum} present many examples of competitions exhibiting intransitivity and argue that it can occur whenever the best strategy in a given comparison depends on the strategy of the opponent, and \cite{smead2019sports} provides a philosophical argument as to why intransitivity is particularly likely to occur in sports. Given this, it is not surprising to find cases of intransitivity in e-sports \citep{makhijani2019parametric, chen2016modeling,duan2017generalized}. Other applications include social choice, real sensory analysis, and election data-sets.

%Intransitivity between pairs of objects can be considered as interaction terms specific to each pair. 
With $n$ competitors there are $n(n-1)/2$ interactions, or intransitivities, so even in round-robin competitions, with $m$ rounds, there are too many terms to estimate efficiently using empirical methods, unless $m/n$ is large.  \cite{causeur20052} proposed an $O(n^2)$ parameter extension of the Bradley-Terry model to address intransitivity. Subsequently $O(nd)$ parametric models have been studied for some fixed $d\in\mathbb{N}$ $(d\ll n)$, see all the models in Section~\ref{sec:lit}, but they lack the flexibility to cover the potentially $O(n^2)$ different intransitivities across $n$ players, leading to bias; whilst they are not parsimonious when the intransitivity is simple, leading to inefficiency.

Although the \textit{concept} of intransitivity is quite clear, there is no established \textit{measure} of the amount of intransitivity in a dataset. In this work, we propose a definition of intransitivity through a distance metric between the assumed probability of paired comparisons under a Bradley-Terry model, and the empirical or model-based probability estimate, such that for any given dataset the magnitude of the intransitivity present is unambiguous. A \textit{flexible} model then, is one which is capable of exploring the space of all possible combinations of intransitivity, as defined by this measure. Any parametric model is restricted to a subset of this space by definition, with this restriction being most obviously revealed when assessing predictive performance.

We then develop a novel semi-parametric extension of the Bradley-Terry model, allocating the $n(n-1)/2$ pairs of objects to a random number $K$, with $0\le K \le n(n-1)/2$, of distinct intransitivity levels, each level representing a different strategy. 
We term this model the {\it Intransitive Clustered Bradley-Terry} (ICBT) model. Relative to the aforementioned parametric models, this ICBT model provides greater flexibility to enable the incorporation of varying structures, and degrees of, intransitivity. As many of these strategies will have similar effects, we anticipate that $K$ should be small, yet the random property of $K$ provides the potential for it to be large when required. This flexibility ensures that our model is parsimonious, whatever the complexity of the data. For our Rock-Paper-Scissors illustration $K=1$. 

Moreover, our novel approach for the objects' skills is to allocate the $n$ objects into a random number of $A+1\leq n$ distinct skill levels, to improve efficiency and avoid false rankings. This constraint recognises that from paired comparison data there will be objects that are indistinguishable as having statistically significantly different skill levels, e.g., for our Rock-Paper-Scissors illustration $A=0$. So clustering skills avoids over-interpretation of misinformed rankings, a feature \cite{masarotto2012ranking} address by clustering skills via a lasso procedure.

The basis of our model is the belief that in practice there are likely to small subsets of skill and intransitivity levels, namely $A\leq n-1$ and $K \ll n(n-1)/2$ respectively.
As we have little prior knowledge about the skills of the objects or the intransitivities of the pairs of objects, we allow the clustering of objects into different skill levels, and of the pairs of objects into separate intransitivity levels, to be determined entirely through a Bayesian hierarchical model. 
We take each of $(A, K)$, the allocations of objects to skill levels, and the allocations of the pairs of objects to intransitivity levels as unknown, with inference being conducted via a reversible jump Markov chain Monte Carlo (RJMCMC) algorithm. This formulation does offer computational challenges; however, we anticipate that typically the posterior will give a high probability that $A+K<n$ and that many of the cluster allocations also will be strongly identified. Our inference framework offers the opportunity to select a highly simplified model, with the values of $A, K$ and allocations fixed at values given by posterior means/modes if these are found to align with known structure about the paired comparison. In the absence of such knowledge our results allow for the full uncertainty of these features to be accounted for.

In certain circumstances our model has the potential to identify and correct for imbalanced tournament structure on overall rankings since teams are not penalised if they (unfairly) compete most frequently against those whom they perform systematically worse to relative to what is expected based on respective skills alone.

We use American League Baseball data to illustrate the performance of our methods in comparison to existing models for a range of reasons. Firstly, each game results in a win or a loss for a team. Secondly, it is known to be a highly strategic game, see Section \ref{sec:baseball}, so we anticipate that the level of intransitivity will be high. Finally, although the tournament structure is not round robin, each team plays each other team often, and so the existence of intransitivity  
%so when coupled with the multiple occurrence of pairings any intransitivity 
should become apparent in inference. Indeed this is found in Section~\ref{sec:baseball}, where our model is shown to have an improved fit over the Bradley-Terry model and existing parametric intransitivity models in out of sample testing for each of the nine seasons we study.

The layout is as follows. Section \ref{sec:lit} introduces other approaches to modelling intransitivity. Section \ref{sec:model} then introduces our novel measure of intransitivity, the ICBT model, and the ranking formulation. Section \ref{sec:inf} contains details of the inference, including prior specification, our full Bayesian hierarchical modelling strategy, an overview of the RJMCMC algorithm and its novel features, and an overview of a simulation study. Section \ref{sec:baseball} compares this model with the Bradley-Terry model and other competitor models, using American League baseball data. Section \ref{sec:Concl}  is a discussion. Full details of the RJMCMC algorithm, simulation study, and extended analysis of the baseball application are in the supplementary material.

\section{Literature on Intransitive Modelling}
\label{sec:lit}
\label{sec:litBT}
The blade-chest model of \cite{chen2016modeling}, extends the Bradley-Terry model into $d$-dimensions by incorporating so-called \textit{blade} and \textit{chest} vectors $b_i, c_i \in \mathbb{R}^d$ for each object $i\in \mathcal{I}$. 
There are two versions: the \textit{-dist} and \textit{-inner} variants, given respectively by 
$$\text{logit}\left(p^{(BCD)}_{ij}\right) := ||b_j -  c_i||_2^2 - ||b_i - c_j||_2^2 + r_i - r_j,\mbox{ and }\text{logit}\left(p^{(BCI)}_{ij}\right) := b_i^T\cdot c_j - b_j^T\cdot c_i + r_i - r_j.$$
The blade and chest parameters of all the objects can be viewed as features on a $d$-dimensional latent space. Then, if an object $i$'s blade is close to object $j$'s chest, and simultaneously object $i$'s chest is far from object $j$'s blade, then object $i$ has an additional advantage over object $j$. If $d=2$ this model can represent a deterministic Rock-Paper-Scissors game, by placing the blade of Rock at the chest of Scissors, the blade of Scissors at the chest of Paper, and the blade of Paper at the chest of Rock. By increasing $d$, ever more complex relationships can be captured between the pairs of objects. Given $n$ objects and $b_i,c_i \in \mathbb{R}^d$ for each object $i$, the model contains $2d(n-1)$ identifiable parameters. The $r$ parameters can be absorbed into the blade and chest parameters; however, the above parametrisation makes it clear that the Bradley-Terry model is a special case of the blade-chest model, when $b_i = b_j = c_i = c_j ,\;\forall i,j \in \mathcal{I}$.
 
\cite{duan2017generalized} introduce a generalised model for intransitivity, with $$\text{logit} \left(p_{ij}^{(G)}\right) = \mu_i^T\Sigma \mu_j + \mu_i^T\Gamma \mu_i - \mu_j^T\Gamma \mu_j,$$ with $\mu_i \in \mathbb{R}^{d}$, where $d$ is even, being a $d$-dimensional \textit{strength} vector, for an object $i \in \mathcal{I}$, and ${\Sigma,\Gamma \in \mathbb{R}^{d\times d}}$ are so-called \textit{transitive matrices}. The first matrix $\Sigma$ represents the interactions between objects, and $\Gamma$ controls how an individual object's strength components form the object's overall strength. The number of identifiable parameters is $d(3d/2 + n-1)$, since there are two $d\times d$ matrices $(\Sigma,\Gamma)$, of which $\Sigma$ is skew-symmetric, and $n$ $d$-dimensional vectors. They show that their model is a generalisation of the blade-chest and Bradley-Terry models. Specifically, the blade-chest-inner model arises when $\mu_i = (b_i,c_i),$ $\mu_j = (b_j,c_j)$, and $||\mu_i||_2^2 = ||\mu_j||_2^2$, that is, the objects all have equal skill, and $\Sigma$ is a block diagonal matrix with two $(d/2)\times (d/2)$ matrices of zeros on the diagonal and matrices $I_{d/2}$ and $-I_{d/2}$ on the off-diagonals where $I_m$ is the $m\times m$ identity matrix, then $\mu_i^T\Sigma \mu_j =  b_i^T\cdot c_j - b_j^T\cdot c_i.$ 
The degrees of freedom are restricted by regularization, using an $L_2$ norm for the object strength vectors and Frobenius norm for both transitivity matrices. The tuning parameters 
%which control the degree of regularisation 
are selected via cross-validation.% which can be computationally challenging.

\cite{makhijani2019parametric} introduced a majority vote model with object $i$ having a vector of $d$ skill attributes, $\left( \mu_{i,1},\dots,\mu_{i,d}\right)$, where $d$ is odd. Then, given a suitable choice of mapping function $f$, e.g., logistic or Gaussian, define $q^l_{ij} = f(\mu_{i,l} - \mu_{j,l}),$ ${\;\forall l \in \{1,\dots,d\}}$ to be the probability of $i$ beating $j$ based only on their $l$th attribute. Then, majority vote model says that the probability of $i$ being preferred to $j$ overall, is the probability that it wins across the majority of attributes. 
For $d=1$ the model is linearly transitive, but not when $d=3$, as $$\Pr\{i \succ j \} = q^1_{ij}q^2_{ij}q^3_{ij} + (1-q^1_{ij})q^2_{ij}q^3_{ij} + q^1_{ij}(1-q^2_{ij})q^3_{ij} + q^1_{ij}q^2_{ij}(1-q^3_{ij}).$$ 
\section{Modelling}
\label{sec:model}
\subsection{Measure of Intransitivity}
From the model definition \eqref{eq:BT_definition}, the Bradley-Terry model assumes linear transitivity. This assumption constrains the pairwise probabilities of the model such that, given $p^{(\text{BT})}_{ij}$ and $p^{(\text{BT})}_{jk}$ from \eqref{eq:BT_definition} for any $i\neq j \neq k \in \mathcal{I}$, the probability $p^{(\text{BT})}_{ik}$ is completely determined. It is straightforward to show that the form of $p^{(\text{BT})}_{ik}$ is given as %P_T\left(p^{(\text{BT})}_{ij}, p^{(\text{BT})}_{jk}\right) :=
 $$p^{(\text{BT})}_{ik} = \frac{p^{(\text{BT})}_{ij}p^{(\text{BT})}_{jk}}{1 + 2p^{(\text{BT})}_{ij}p^{(\text{BT})}_{jk} - \left(p^{(\text{BT})}_{ij} + p^{(\text{BT})}_{jk}\right)},\; \forall j \neq i,k,$$ noting it is independent of the choice of \textit{bridge} object $j$. Therefore, there can be no interaction that is specific to the pair $\{i,k\}$, that is not already captured between all other pairs.
 %$\{i,j\}, \{j,k\},\;\forall j\neq i,k \in \mathcal{I}$. 

Including intransitivity; however, allows for some pairwise probabilities to depart from those assumed by the Bradley-Terry model. This can be formalised by supposing that for all $i\neq k \in \mathcal{I}$ the true probability of preference $i \succ k$ is given as some function ${f:\{[0,1], \mathbb{R}\}\rightarrow [0,1]}$ of the Bradley-Terry probability and the intransitivity, $\theta_{ik}$, of the pair $\{i,k\}$, then we can write
\begin{equation}
\label{eq:general_IBT}
p_{ik} := f\left(p^{(\text{BT})}_{ik}, \theta_{ik} \right),\;\forall i\neq k \in\mathcal{I},
\end{equation}
where we identify the form of $f$ in Section~\ref{sec:model_def}. We define the intransitivity to be the displacement of the true probabilities from the Bradley-Terry probabilities on the log-odds scale, so that
 \begin{equation}
\label{eq:theta}
\theta_{ik} := \log\left(\frac{p_{ik}/(1-p_{ik})}{p^{(BT)}_{ik}/(1-p^{(BT)}_{ik})}\right),\;\forall i \neq k \in \mathcal{I},
\end{equation}
is the amount of intransitivity between the pair of objects $\{i,k\}$. A value of $\theta_{ik}=0$ indicates that the comparison is transitive, i.e., the pairwise probabilities could be modelled by the Bradley-Terry model. As a consequence we require $f(x,0) = x,\; x \in [0,1]$. The choice of log-odds ratio in equation \eqref{eq:theta} reflects the non-linearity of probabilities. For example, if $\epsilon=0.099$, then a small linear shift in probability from $0.5$ to $0.5+\epsilon$ has little impact on the odds, which remain at approximately 1:2. However, a linear shift in probability from $0.9$ to $0.9+\epsilon$ has a huge impact on the odds, which move from 1:10 to 1:1000. Moreover, our definition \eqref{eq:theta} for the intransitivity $\theta_{ik}$ also imposes rotational symmetry for pairs of objects, that is \begin{equation}
\label{eq:symmetry}
\theta_{ki} = \log\left(\frac{p_{ki}/(1-p_{ki})}{p^{(BT)}_{ki}/(1-p^{(BT)}_{ki})}\right) = \log\left(\frac{(1-p_{ik})/p_{ik}}{(1-p^{(BT)}_{ik})/p^{(BT)}_{ik}}\right) =- \theta_{ik},\;\forall i \neq k \in \mathcal{I},
\end{equation} so we need to find $\{\theta_{ik},\;\forall i>k \in \mathcal{I}\}$ only, in order to completely define $\{\theta_{ik},\;\forall i \neq k \in \mathcal{I}\}$.

\subsection{Model formulation}
\label{sec:model_def}

To find the function $f$ in equation \eqref{eq:general_IBT}, equation \eqref{eq:theta} can be simply rearranged which %, using the new definition of $\theta_{ik}$ from equation \eqref{eq:thetanew}, 
gives \begin{equation}
\label{eq:IBTmodel}
p_{ik} = \frac{p^{(BT)}_{ik} \exp(\theta_{ik})}{p^{(BT)}_{ik}\exp(\theta_{ik}) + 1-p^{(BT)}_{ik}}, \; \forall i\neq k \in \mathcal{I},
\end{equation}
and so for any pair $\{i,k\}$, equation \eqref{eq:IBTmodel} can be re-written as \begin{equation}
\label{eq:IBT2}
p_{ik} = \frac{1}{1 + \exp[-\left(\theta_{ik} +r_{i} - r_{k}\right)]},\;\forall i \neq k \in \mathcal{I}.
\end{equation}
Here the effect of $\theta_{ik}$ is clear,  positive (negative) $\theta_{ik}$, increases (decreases) the probability of team $i$ beating team $k$ relative to their skills alone, i.e., relative to the Bradley-Terry model.

Thus far, the model contains the flexibility to describe ${P := \{p_{ik} \in [0,1],\forall i \neq k \in \mathcal{I}\}}$ completely. Here $P$ contains $n(n-1)/2$ degrees of freedom, because $p_{ki} = 1-p_{ik}$; however, the model contains $n(n-1)/2 + n$ parameters: $n(n-1)/2$ values of intransitivity between pairs, and $n$ skill parameters from $r$, and thus the model parameters are not identifiable. One way of ensuring identifiability in the standard Bradley-Terry model is to fix one object's skill level, and here it is chosen that $r_1=0$. As well as this constraint on the objects' skill parameters, the intransitivity parameters require constraints for parameter identifiability. The minimal set of required constraints is identified in Proposition \ref{prop:ID}, see the Appendix for the proof.
\begin{prop}
\label{prop:ID}
Consider a round-robin tournament with pairs of objects $(i,j)$ being compared, with $i,j\in \mathcal{I}$, with $i\neq j$ where $|\mathcal{I}|=n$. Suppose that the probabilities of $i$ beating $j$ are $p_{ij}$ where these probabilities are given by expression \eqref{eq:IBT2}, with $r_1 = 0$ and intransitivity values $\theta_{ij}$. If a set of $n-1$ pairs of objects, indexed by $\mathcal{J}_{n-1}$, have their intransitivity values set to arbitrary specified values, then all the rest of the $\{r_i\}$ and $\{\theta_{ij}\}$ parameters in expression \eqref{eq:IBT2} are identifiable if $\mathcal{J}_{n-1}$ forms a connected graph over $\mathcal{I}$. Furthermore, if less than $n-1$ pairs' intransitivity values are specified or if 
$\mathcal{J}_{n-1}$ is not a connected graph over $\mathcal{I}$ then identifiability is not achievable.
\end{prop} 
We choose the $n-1$ constraints to be $\theta_{ij} = 0 ,\;\forall (i,j):i = 1, j \in \mathcal{I}\setminus \{1\}$, that is, all pairs involving object 1 have intransitivity set to 0. Proposition~\ref{prop:ID} gives that if any further constraints are imposed on the intransitivity values the flexibility of model~\eqref{eq:IBT2} will be compromised.

With the above constraints, the minimal conditions for parameter identifiability are satisfied, but the model is still likely to overfit with so many parameters. To rectify this we restrict the total number of degrees of freedom, by restricting both the number of intransitivity values to only $K \leq n(n-1)/2$ unique values and restricting the number of unknown skill values to be $A < n$,
where $A + K \leq n(n-1)/2$ and ideally $A + K \ll n(n-1)/2.$ In this fashion our ICBT model embraces intransitivity in a parsimonious way. %, with the BT model corresponding to $K = 0$. 

Firstly, consider the $A+1$ unique \textit{skill values}, which ensures parsimony in the model 
%The Clustered Intransitive Bradley-Terry model, as well as clustering the degree of intransitivity, is made parsimonious 
by clustering the objects' skills $r$ into distinct values which are sufficiently statistically significantly different. Since $r_1=0$ is fixed, there are only $A$ unknown \textit{skill levels}, $\phi\in \mathbb{R}^A$. By defining the labels of the set of skill levels to be ${\mathcal{A} := \{-A_- ,\dots,0,\dots,A_+\}}$ with $A_+$ being the number of skill levels which are greater than 0 and $A_-$ the number of skill levels less than 0 such that $A_- + A_+ + 1= A+1 = |\mathcal{A}| \leq n$, we impose the equivalent condition in our model by fixing the skill level with label $\{0\}$ to be $\phi_0 = 0$, and fixing object 1 to always be allocated to this cluster.
The possible skill values an object can take are therefore defined as $$\{\phi_0 = 0, \phi := \{\phi_a \in \mathbb{R},  \forall a \in \mathcal{A}\setminus\{0\}\} :\phi_{-A_-} < \dots < \phi_0 < \dots < \phi_{A_+}\},$$ where $\phi$ are the unknown \textit{skill levels}, and the ordering helps with label switching problems in the inference. The \textit{skill cluster allocation} of object $i$, denoted $y_{\{i\}} \in \{0,1\}^{A+1}$, is a binary vector which takes the value 1 at position $s\in \mathcal{A}$ and 0 everywhere else, if object $i\in\mathcal{I}$ belongs to cluster $s$.
%, see \cite{spearing2022Supplementarybaseball} for a formal definition. 
The set ${Y} := \{{y}_{\{i\}} : i \in \mathcal{I}\setminus\{1\}\}$ then contains all the objects' skill cluster allocations except object $\{1\}$ which has fixed cluster allocation. Therefore, by defining $\mathcal{S}_{\{i\}}\left(Y\right) := \text{argmax}_s y_{\{i\},s},\;\forall i\in \mathcal{I}\setminus\{1\}$, then the objects' skills can be written as \begin{equation}
\label{eq:stregth_def}
r_i = \left\{\begin{matrix}
\phi_{\mathcal{S}_{\{i\}}\left(Y\right)}, & i \in \mathcal{I}\setminus\{1\}\\
0 , & i=1
\end{matrix}\right.\quad
:= f_r\left(\phi,Y,i\right),\;  \forall i \in \mathcal{I}.
\end{equation}

Now consider the $K$ unique values of intransitivity to describe the different inter-object strategies. Of the $n(n-1)/2$ pairs of objects, many will adopt similar strategies depending on their opponents. These similar strategies are translated by the model as having similar departures from transitivity, and are thus clustered together. For example, suppose some group of objects $\mathcal{V}: j \notin \mathcal{V}$ competed against object $j$ in the same way. Then it would be reasonable to assume that $\theta_{ij}$ is the same for all $i \in \mathcal{V}$. This creates clusters of pairs of objects, such that the pairs are clustered according to them having identical intransitivity.

In order to measure the departure from a Bradley-Terry model, a \textit{linearly transitive cluster} is imposed, which contains the set of pairs ${\mathcal{J}_T \subseteq \{\{i,k\}:i \neq k \in \mathcal{I}\}},$ which have an intransitivity level $\theta_0 = 0$. Thus, the Bradley-Terry modelling assumption \eqref{eq:BT_definition} holds for these pairs, such that ${p_{ik} = p^{(\text{BT})}_{ik}}$, for all $\{i,k\} \in \mathcal{J}_T$. Given the existence of this cluster, there must be strong evidence from the data to produce an additional cluster with an intransitivity level close to 0. This choice does not impose transitivity of pairs as the linearly transitive cluster $\mathcal{J}_T$ may be empty, except for all pairs with object 1 which are classified as transitive due to our imposition of constraints for identifiability from Proposition~\ref{prop:ID}.
Let the distinct set of intransivity levels be $$\theta_\mathcal{K} := \{\theta_0=0,\;\theta = \{\theta_k\in\mathbb{R}_+,\;\forall k \in \mathcal{K}\}: 0 < \theta_1 < \dots < \theta_K\},$$ where $\mathcal{K} = \{1,\dots,K\}$ and the levels of intransitivity are ordered from smallest to largest. The \textit{levels of intransitivity}, $\theta$, contain the set of positive values of intransitivity which, due to symmetry and the completely transitive cluster with intransitivity value $\theta_0$, then define the full $2K +1$ possible values of intransitivity between any pair of objects.

We define the intransitivity cluster allocation of a given pair $\{i,k\}$ to be another binary matrix $z_{\{i,k\}}$, which takes the value 1 at position $s \in \{-K,\dots,K\}$ and 0 everywhere else, if the pair $\{i,k\}$ belongs to cluster $s$.
%see \cite{spearing2022Supplementarybaseball} for a formal definition. 
The clusters are therefore labelled from $-K$ to $K$, where a cluster labelled $k \in \{1,\dots K\}$ has cluster level $\theta_k$, a cluster labelled $k \in \{-K,\dots,-1\}$ has cluster level $-\theta_{-k}$, and a cluster with label $0$ has cluster level $\theta_0=0$. The set $Z := \{z_{\{i,k\}},\;\forall i>k \in \mathcal{I}\setminus \{1\}\}$ then defines all the cluster allocations for all the free pairs $i \neq k \in \mathcal{I}\setminus \{1\}$, because of the rotational symmetry. For example, if the $K$th index of $z_{\{i,k\}}$ has value $z_{\{i,k\},K} =1$, then this indicates that the pair $\{i,k\}$ belongs to the cluster with label $K$, whose cluster level is the largest level of intransitivity $\theta_K$, and this enforces that the pair $\{k,i\}$ belongs to cluster $-K$ and has the smallest level of intransitivity $-\theta_K$. If the cluster allocation of the pair $\{i,k\}\in \mathcal{I}\setminus \{1\}$
is
\begin{equation}
 \label{eq:S_alloc}
 \mathcal{S}_{\{i,k\}}(Z) :=\left\{ \begin{matrix} \text{argmax}_s z_{\{i,k\},s} & \text{if } i>k,\\
  -\text{argmax}_s z_{\{k,i\},s}& \text{if } i<k,\\
  \end{matrix}\right.
 \end{equation}  
 then the level of intransitivity for a pair $\{i,k\}$, $\theta_{ik}$ can be redefined as \begin{equation}
\label{eq:thetanew}
\theta_{ik} = \left\{\begin{matrix}
 \theta_{\mathcal{S}_{\{i,k\}}(Z)}\mathbbm{1}\{\mathcal{S}_{\{i,k\}}(Z)\geq 0\} - \theta_{-\mathcal{S}_{\{i,k\}}(Z)}\mathbbm{1}\{\mathcal{S}_{\{i,k\}}(Z)<0\}, & \{i,k\}\in \mathcal{I}\setminus \{1\}\\
 0,& \text{otherwise}
 \end{matrix}\right.\quad  := f_\theta\left(\theta, Z, \{i,k\}\right)
\end{equation} where $\mathbbm{1}$ is the indicator function, and remembering that $\theta_0 = 0$.% This now has completely defined the Intransitive Bradley-Terry model.

The full model can be written either in terms of equation \eqref{eq:IBT2}, noting that the parameters will be clustered, or can be written in terms of the levels and the cluster allocations, 
\begin{equation}
\label{eq:IBT3}
p_{ik} = (1 + \exp\left\{-\left[f_\theta\left(\theta,Z,\{i,k\}\right) + f_r\left(\phi,Y,i\right) - f_r\left(\phi,Y,k\right)\right]\right\})^{-1}.
\end{equation}
So the ICBT model is defined by ${\psi =\{ \phi = \{\phi_a: a\in\mathcal{A}\setminus\{0\} \},\; \theta = \{\theta_k:k \in \mathcal{K}\} \}}$.

Due to the intransitivity levels being fixed to 0 for all pairs of objects involving object 1, an adjustment is required to get a more interpretable value of intransitivity between the pairs. We define the adjusted intransitivity to be \begin{equation}
\label{eq:intrans2}
\theta^*_{ij} := \text{logit}\left(p_{ij}\right) - \text{logit}\left(p^{(BT)}_{ij}\right) = \theta_{ij} + r_i - r_j - \left( r^{(BT)}_i - r^{(BT)}_j\right),
\end{equation}	
that is, the difference between the logits of the pairwise probability between our ICBT model and the Bradley-Terry model. Note that the rotational symmetry of $\{\theta_{ij}\}$ \eqref{eq:symmetry} also imposes rotational symmetry on $\{\theta^*_{ij}\}$, that is, $\theta^*_{ij} =\theta^*_{ji}, \forall i\neq j \in \mathcal{I}$.

To help see the value of this reparametrisation, consider then the earlier example of a deterministic game of Rock-Paper-Scissors.  Take Rock as the constrained object, then Rock has fixed skill level $r_r = 0$, and that pairs involving Rock have intransitivity $0$, that is $\theta_{rp}=\theta_{rs}=0$, where the $p$ and $s$ subscripts denote Paper and Scissors. To maintain that Rock always beats Scissors $p_{rs}=1$, then from the constraints, we get an excellent approximation from the ICBT model when $r_s=-M$ for some large $M$, with the approximation improving as $M\rightarrow \infty$. Likewise $r_p=M$, and $\theta_{ps}=-3M$. With this model there is only one skill level $M$, and one non-zero intransitivity level $-3M$. This parametrisation somewhat hides the symmetry of the intransitivity over pairs. However, with definition~ \eqref{eq:intrans2}, then $\theta^*_{rs} = \theta^*_{sp} = \theta^*_{pr} = M$, resulting in an intuitive and easy interpretation of the intransitivity, reflecting the symmetry of the game, no-matter the choice of the fixed parameters.

\subsection{Model Ranking}
\label{sec:model_interp}
In the Bradley-Terry model, the skill parameters can simply be ordered to give a rank since a greater skill always results in higher win probabilities against all other objects. In our ICBT model this is not the case, because both the intransitivity parameters of each pair and the skill parameters of the objects impact the win probability between any pair. However, below we present two intuitive methods for determining overall ability, and therefore ranking.

Firstly, if $p_{ij} = \Pr\{i\succ j\}$ is the probability of an object $i$ beating object $j$ according to our model, then we can rank the objects by ordering \begin{equation}
\label{eq:ranking_p}
{p}_. :=  \left\{p_{i.} := \frac{1}{n-1}\sum_{j \in \mathcal{I}:\;j\neq i} p_{ij}:i \in \mathcal{I}\right\},
\end{equation}
that is, $p_{i.}$ is the average probability of object $i$ beating any other object $j\neq i \in\mathcal{I}$.

Secondly, if we consider the intransitivity between an object $i$ and an opposing object $j\neq i$ as some ``boost'' which contributes to the overall ability (which could be negative), then the overall ability $a_i$ of object $i$ could be defined by \begin{equation}
\label{eq:ranking_a}
a_i := r_i + \frac{1}{n}\sum_{j \in \mathcal{I}} \theta_{ij},\;\text{where}\; \theta_{ii} = 0,\; \forall i \in \mathcal{I},
\end{equation}
that is, the object skill plus its average intransitivity level. Definition \eqref{eq:ranking_a} is equivalent to the Bradley-Terry definition of `ability'. Defining $\text{logit}\left(p_{ii}^{(BT)}\right)=0,\; \forall i\in\mathcal{I}$, a Bradley-Terry gives \begin{equation}
\label{eq:equatebt}
\frac{1}{n}\sum_{j \in \mathcal{I}}\text{logit}\left(p^{(BT)}_{ij}\right) = r_i - \frac{1}{n}\sum_{j \in \mathcal{I}}r_j,
\end{equation}
where the sum on the right hand side does not depend on $i$, so the skill of object $i$ is entirely determined by $r_i$. Similarly, in our model \begin{equation}
\label{eq:equateicbt}
\frac{1}{n}\sum_{j \in \mathcal{I}}\text{logit}\left(p_{ij}\right) = r_i + \frac{1}{n}\sum_{j \in \mathcal{I}}\theta_{ij} - \frac{1}{n}\sum_{j \in \mathcal{I}}r_j = a_i -\frac{1}{n}\sum_{j \in \mathcal{I}}r_j,
\end{equation} 
then given definition \eqref{eq:ranking_a}, both \eqref{eq:equatebt} and \eqref{eq:equateicbt} have the same form but with $a_i$ replacing $r_i$. Then a ranking can be formed by ordering the set of abilities ${a} := \{a_i:i\in\mathcal{I}\}$. We argue that the first method, using the probabilities ${p}_.$ to rank the objects, is more meaningful since it is directly associated with the pairwise probabilities, the modelling of which is our ultimate aim. The application to baseball data of both methods is discussed in the supplementary material.

\section{Inference}
\label{sec:inf}

\subsection{Likelihood}
The data, $x := \{x_c:c \in \mathcal{C}\}$, are binary, and $i \succ j$ denotes that $i$ is preferred to $j$. Then $x_c =1$ if $i_c \succ j_c$, and $x_c =0$ otherwise, where $i_c, j_c \in \mathcal{I}$ are the objects being compared in comparison $c$. %Furthermore, let $\pmb{z} = \mathbb{R}^{n\times n}$, represent the cluster allocation with $(i,k)$-th entry $z_{ik} \in \{1,\dots,m\}$, where $m$ is odd.
Then, the log likelihood for the ICBT model is
\begin{equation}
\label{eq:IBT_likelihood}
\ell({x}|{\phi}, {Y},A,{\theta}, {Z},K) = \sum_{c \in \mathcal{C}}\left[ x_c \log\left(p_{i_cj_c}\right) + (1-x_c)\log\left(1-p_{i_cj_c}\right)\right],
\end{equation}
where $p_{i_cj_c}$ is given by the ICBT model for all $c \in \mathcal{C}$ and is calculated from the set of parameters $({\phi}, {Y},A,{\theta}, {Z},K)$. All pairs' intransitivities ${\{\theta_{ik}: i\neq j \in \mathcal{I}\}}$ can be found from the intransitivity levels ${\theta}$ and the cluster allocations ${Z}$, using equation~\eqref{eq:thetanew}, so it is only necessary to do inference on these parameters, rather than the full $2K+1$ separate clusters. Therefore from here onwards the term \textit{intransitivity levels} refers only to those $K$ values which have positive intransitivity. Similarly, any individual object's skill $r_i\;\forall i \in \mathcal{I}$ can be found from knowing the ability levels ${\phi}$ and the cluster allocations ${Y}$, using equation \eqref{eq:stregth_def}. We formulate a Bayesian hierarchical model, which treats both $K$ and $A$ as unknown parameters, thus accounting for uncertainty in the number of clusters. The posterior is therefore written as $$\pi\left({\phi}, {Y},A,{\theta},{Z},K|{x}\right) \propto L\left({x}|{\phi}, {Y}, A,{\theta},{Z},K\right) \pi\left({\phi}, {Y},A,{\theta},{Z},K\right)$$ where $L(\cdot) = \exp[\ell(\cdot)]$ is the likelihood and $\pi\left({\phi}, {Y},A,{\theta},{Z},K\right)$ is the prior. 
\subsection{Prior Specification}
\label{sec:prior_spec}
Formulating the prior, we make the assumption that ${Z} \independent {\theta} |K$ that is, the intransitivity level allocations and intransitivity levels are independent from one another given the number of intransitivity levels $K$. Likewise, it is assumed that ${Y} \independent {\phi} |A$. Furthermore, we assume that the clustering of the objects' skills and the clustering of the pairs' intransitivities are independent systems, that is, $A \independent K$, ${\phi} \independent {\theta}$, and ${Y} \independent {Z}$. This means that the prior specification for the two features we are clustering, skills and intransitivities, can be approached separately.

Consider first the prior specification for the clustering of the intransitivity values of the pairs. Remember that labels ${z}_{\{i,j\}}, \forall i >j \in \mathcal{I}\setminus \{1\}$ have domain $\{-K,\dots,K\}$, that is, ${{z}_{\{i,j\}}:\{-K,\dots,K\} \rightarrow \{0,1\},\; \forall i >j \in \mathcal{I}\setminus \{1\}}$, and also that ${z}_{\{i,1\}}, \forall i \in \mathcal{I}\setminus \{1\}$ (and by symmetry ${z}_{\{1,j\}}, \forall j \in \mathcal{I}\setminus \{1\}$ too) are fixed in the transitive level $\{0\}$ for identifiability purposes, see Section \ref{sec:model_def}. Let the prior on the cluster allocation be $${z}_{\{i,j\}}|{\omega}_K \sim \text{multinomial}\left(1,{\omega}_K\right),\; \forall i>j \in\mathcal{I}\setminus\{1\},$$
where ${\omega}_K$ is on $\{-K,\dots,K\}$ such that 
$${\omega}_K = \{\omega_{K,s} \in [0,1]: s \in \{-K,\dots,K\},\; \sum_{s = -K}^{K} \omega_{K,s} = 1 \}.$$ 

The distribution of ${Z}|\left({\omega}_K,K\right)$ is assumed independent over all pairs $i>j \in\mathcal{I}\setminus\{1\}$, i.e., 
%$$f\left(\pmb{Z}|\pmb{\omega}_K,K\right) = \prod_{i>j \in \mathcal{I}\setminus\{1\}}f\left(\pmb{z}_{\{i,j\}}|\pmb{\omega}_K,K\right)= \prod_{s=-K}^K \omega_{K,s}^{|\pmb{b}_s|},$$
$$f\left({Z}|{\omega}_K,K\right) = \frac{\left( \sum_{k = -K}^K |{b}_k|\right)!}{\prod_{k= -K}^K |{b}_k|!}\prod_{s=-K}^K \omega_{K,s}^{|{b}_s|},$$
where ${b}_k = \{ (i,j):i>j \in\mathcal{I}\setminus\{1\}:z_{\{i,j\},k}=1\}\;\forall k \in \{-K,\dots,K\}$ is the set of allocated pairs of objects belonging to cluster $k$. We set ${{\omega}_K|K \sim \text{Dirichlet}\left(\bar{\gamma}_K\right)}$ to come from $2K+1$ dimensional Dirichlet prior distribution, and $\bar{\gamma}_K \in \mathbb{R}_+^{2K+1}$ is the hyper-parameters vector. We use an uninformative prior, setting $\bar{\gamma}_K = \gamma_K {1}_{2K+1}$ where ${1}_{2K+1}$ is a vector of ones of length $2K + 1$ and $\gamma_K \in \mathbb{R}_+$. In this case, the ${\omega}_K$ parameter can be marginalised out, by
\begin{align}
\label{eq:priorAlloc}
f({Z}|\gamma_K,K) =& \int_{{\omega}_K} f({Z}|{\omega}_K,K)f({\omega}_K|\gamma_K,K) \diff {\omega}_K\nonumber\\
=& \frac{\left( \sum_{k = -K}^K |{b}_k|\right)!}{\prod_{k= -K}^K |{b}_k|!}\frac{\Gamma((2K+1)\gamma_K)}{\Gamma(\gamma_K)^{2K+1}}\frac{\prod_{k= -K}^K \Gamma\left( \gamma_K + |{b}_k|\right)}{\Gamma\left((2K+1)\gamma_K + \sum_{k = -K}^K |{b}_k|\right)}
\end{align}
where integration on ${\omega}_K$ is taken over the $2K+1$ simplex. This is referred to as a Dirichlet-multinomial allocation prior. The prior for $K$ is a Poisson$(\lambda_K)$ distribution with probability mass function denoted $g_0\left(k|\lambda_K\right)$ % = \frac{\lambda_K^{k} \exp\left(-\lambda_K\right)}{k!}: k = 0,1,2,\dots,$$
so that $\mathbb{E}\left[K|\lambda_K\right] =\lambda_K$, with $\lambda_K > 0$. Note that $K = 0$ is feasible, as this corresponds to the Bradley-Terry model since ${\theta}|(K=0) = \emptyset$ and so only the transitive cluster exists, that is ${\theta}_\mathcal{K}|(K=0) = \theta_0$, and $\{i,j\} \in \mathcal{J}_T,\;\forall i \neq j \in \mathcal{I}$, so all pairs belong to the transitive cluster $\mathcal{J}_T$. Formally $K<n(n-1)/2-n$ but as this is large relative to our prior beliefs on $K$, for simplicity we ignore this constraint in the inference.

As the ${\theta}$ elements are ordered in increasing order and are positive, the prior on the ${\theta}$ parameters is taken to be the joint distribution of $K$ order statistics drawn from independent gamma random variables, such that \begin{equation}
\label{eq:prior_intrans_single_cluster}
h_0\left({\theta}|K\right) = K!\prod_{i = 1}^K h_0\left(\theta_i|\alpha, \beta\right),\; \text{with } 0 < \theta_1 < \dots < \theta_K,\; K\geq 1,\end{equation}
and where $h_0\left(x|\alpha, \beta\right)$ is the Gamma$(\alpha, \beta)$ density
%= \frac{\beta^\alpha}{\Gamma(\alpha)}x^{\alpha-1}\exp\left(-\beta x\right),\; x>0,$ 
with shape and scale $\alpha, \beta > 0$ respectively.

Consider the prior for the  skill levels clustering. The set of skill cluster allocations has distribution
${Y} = \{{y}_{\{i\}}|{\omega}_A, A \sim \text{multinomial}(1, {\omega}_A), \forall i \in \{2,\dots,n\}\},$
where ${\omega}_A$ has domain on $\{-A_- ,\dots, A_+\}$. %and define $${{y}_{\{1\}} := \{ y_t \in [0,1], \; t\in\mathcal{A}: y_0 = 1, y_t = 0:t\neq 0\}},$$ which simply indicates that object $1$ is fixed in the 0 level. 
The distribution of ${y}_{\{i\}}|({\omega}_A,A)$ is assumed independent over all objects $i \in \mathcal{I}\setminus\{1\}$ such that $$f({Y}|{\omega}_A,A) = \prod_{i \in \mathcal{I}\setminus\{1\}} f({y}_{\{i\}}|{\omega}_A,A),$$ 
where 
$${\omega}_A = \{\omega_{A,s} \in [0,1]:s \in\{-A_-,\dots,A_+\},\sum_{s=-A_-}^{A_+} \omega_{A,s} = 1\}.$$
Again, ${\omega}_A|(A,\gamma_A) \sim \text{Dirichlet}\left(\bar{\gamma}_A\right)$ is modelled to come from an $A+1$ dimensional Dirichlet prior distribution, with $\bar{\gamma}_A = \gamma_A {1}_{A+1}$ where $\gamma_A \in \mathbb{R}_+$. Marginalising out as in derivation \eqref{eq:priorAlloc}, another Dirichlet-multinomial allocation prior is obtained by integrating ${\omega}_A$ over the $A+1$ dimensional simplex. The prior density for the skill allocations is therefore given as
\begin{equation}
\label{eq:priorDMA_A}
f({Y}|\gamma_A,A) = \frac{n!}{\prod_{a=-A_-}^{A_+} |{c}_a|!}\frac{\Gamma(A\gamma_A)}{\Gamma(\gamma_A)^A}\frac{\prod_{a=-A_-}^{A_+}\Gamma(\gamma_A + |{c}_a|)}{\Gamma(A\gamma_A + n)},
\end{equation}
because $\sum_{a=-A_-}^{A_+}|{c}_a| = n$, where ${c}_a := \{i,\;\forall i \in\mathcal{I}\setminus \{1\}:{y}_{\{i\},a}=1\}$ is the set of objects belonging to skill cluster $a \in \mathcal{A}$. 

The prior for the number of unknown skill levels $A$ is taken to be a truncated Poisson distribution with parameter $(\lambda_A)$, $\lambda_A>0$ with probability mass function
$$g_A\left(a|\lambda_A\right) = \frac{\lambda_A^{a}}{a!}\left(\sum_{i=0}^{n-1} \frac{\lambda_A^i}{i!}\right)^{-1} a = 0,1,\dots,n-1.$$
%$$g_0\left(a|\lambda_A\right) = \frac{\frac{\lambda_A^{a} \exp\left(-\lambda_A\right)}{(a)!}}{\exp\left(-\lambda_A\right) \sum_{i=1}^{n} \frac{\lambda}{i!}} a = 0,1,2,\dots,$$
Similarly to ${\theta}$, the prior choice for ${\phi}$ is taken to be the joint distribution of order statistics of independent and identically distributed $A+1$ Gaussian random variables such that $$\pi({\phi}|A) = (A+1)! \prod_{a \in \mathcal{A}\setminus\{0\}} \pi(\phi_a) \text{ for } \phi_{A_-} <\dots<\phi_0<\dots\phi_{A_+},$$ where ${\phi_a \sim \mathcal{N}\left(0, \nu_A^2\right)\;\forall a \in \mathcal{A}\setminus\{0\}}$, and with ${\nu_A \in \mathbb{R}_+}$. The $(A+1)!$ term arises as $\phi_0$ can occur anywhere in the sequence of ${\phi}$.

In summary, the prior $\pi\left({\phi}, {Y}, A,{\theta},{Z},K\right)$ is equal to \begin{align}
\label{eq:priorICBT}
& (A+1)!\left[\prod_{a \in \mathcal{A}_{\{-0\}}} \pi\left(\phi_a\right)\right]f({Y}|\gamma_A,A)g_A\left(A|\lambda_A\right)K! \left[\prod_{i = 1}^{K}
h_0\left(\theta_i|\alpha,\beta\right)\right]f({Z}|\gamma_K,K) \;g_0\left(K|\lambda_K\right),
\end{align}  
where, $\lambda_K, \lambda_A, \gamma_K, \gamma_A, \nu_A$, and $\alpha, \beta$ are the hyper-parameters. \\

\subsection{Reversible jump Markov chain Monte Carlo sampler}
Inference is made via a reversible jump Markov chain Monte Carlo sampler \citep{green1995reversible}, which provides samples from the posterior distribution ${\pi\left({\phi}, {Y}, A,{\theta},{Z},K|{x}\right)}$, that is, the intransitivity and skill levels, the allocations to the these levels, and the number of levels.
Since the  number of skill and intransitivity levels ($A, K$) are assumed to be unknown, the uncertainty in these parameters must be accounted for, thus motivating the use of a reversible jump sampler. In a sense, the reversible jump sampler mixes over models as well as parameters, and thus fully accounts for this uncertainty in the final inference.

The ICBT model is structured to try to favour the Bradley-Terry model as a special case, and this is reflected in our sampler, by explicitly incorporating the completely transitive cluster $\theta_0$ as an ever present cluster, even if no pairs are allocated to this cluster at a given iteration of the sampler. To ensure the skill and intransitivity levels both remain ordered, the updates to these levels occur in a transformed space such that no update can lead to a change in order.

The reversible jump algorithm used is a split-merge sampler \citep{green2001modelling}, which is adapted from the work of \cite{ludkin2020inference}. The sampler comprises three separate moves: a standard Markov chain Monte Carlo Metropolis-Hastings move, which samples parameters ${\phi}$, ${\theta}$, and reallocates clusters ${Y},{Z}$; splitting or merging clusters; and adding or deleting empty clusters. For the construction of the algorithm and how it is implemented, see the supplementary material.

\subsection{Model assessment}
The inferences produced by any model are only meaningful if the model itself is accurate. This accuracy is measured here by how well the model fits out of sample. If $\mathcal{C}$ is the set of total observed pairwise
comparisons, then let $\mathcal{C}_s$ be the set of comparisons on which the model is fitted and $\mathcal{C}_t$ be the set of comparisons on which the model performance is analysed, such that $\mathcal{C}_s \cup \mathcal{C}_t = \mathcal{C}$ and $\mathcal{C}_s \cap \mathcal{C}_t = \emptyset$. We use log-loss 
%as it is a more precise measure than simply the number of correct predictions, which captures nothing about the pairwise probabilities.
$l(x^*)$ of the test dataset $x^*$
%The Log-loss of a dataset $\pmb{x}$, $l(\pmb{x})$ is used to
 to measure model performance, which we take to be the average negative log-likelihood per observation in $x^*$, i.e.,  
 \begin{equation}
 \label{eq:logloss}
 l(x^*) = -\frac{1}{|\mathcal{C}_t|}\sum_{c\in\mathcal{C}_t}\left[x_c\log(\hat{p}_{i_cj_c}) + (1-x_c)\log(1-\hat{p}_{i_cj_c})\right],
 \end{equation}
 % \begin{equation}
 %l(\mathcal{C}_t) = -\frac{1}{|\mathcal{C}_t|}\sum_{(ij)\in\mathcal{C}_t}\log\left(\hat{p}_{ij}\right),
% \end{equation}
where $\hat{p}_{ij}$ is the point estimate of $p_{ij}$ based on the training dataset of comparisons $\mathcal{C}_s$ and $x^* := \{x_c:c \in \mathcal{C}_t\}$ is the set of test data, where the notation is as used in the expression for the likelihood \eqref{eq:IBT_likelihood}.

\subsection{Simulation study}
The model was tested using simulated datasets where the number of objects, the number of round-robin tournaments, and the amount of intransitivity varied between the datasets. The sensitivity of our model to these parameters was then tested by comparing out of sample prediction accuracy with a standard Bradley-Terry model. This provided insights into the amount of data, and the amount of intransitivity, required for our more complex model to outperform the Bradley-Terry model. A full analysis is provided in the supplementary material.

\section{Baseball Data}
\label{sec:baseball}

\subsection{Data}
Baseball was chosen to illustrate the methodology due to the high frequency of games, with accessible data for the American League Baseball obtained from \url{www.retrosheet.org}. The data are from the 2010-2018 seasons, with the 2010-2012 seasons involving 14 teams, and the 2013-2018 seasons involving 15 teams due to the Houston Astros moving from the National League to the American League. We analyse each season's data separately here, and jointly over years in the supplementary material. %\cite{spearing2022Supplementarybaseball}.

The tournament structure is not as simple as the round robin tournament we considered in the simulation study. The American League is split into three divisions based on location: East, Central and West, with five teams in each (since 2013). Within the same division, pairs of teams play each other approximately 20 times, and pairs of teams from different divisions play each other around 5-7 times, as well as any Playoffs and World Series matchups, totalling around 140-160 matches per team every season, depending on the season and the team. 
Baseball is known to be a highly strategic game, with issues such as player selection, handedness of the of batters, strength and speed of players, and tactics such as ``small ball'' vs ``long ball'' all considered of great importance. So we anticipate that the level of intransitivity will be high.

%Several extensions have been made to the Bradley-Terry model to incorporate draws \citep{davidson1970extending, rao1967ties}, and could be incorporated into our ICBT model in a similar fashion, see Section~\ref{sec:Disc}, but here only win-loss situations are considered. 

The vast majority (at least $99.5\%$) of all matches are played at the home of one of the two teams competing in the game, with the rest played at neutral venues. Playing at home is well known to have the potential to increase the probability of the home team winning the match across a range of sports \citep{dixon1997modelling}. Although prediction and model interpretation could be improved by incorporating this effect, we decided not to address home advantage here. Our reason was that none of the existing intransitivity models have such a feature, as they were developed for applications devoid of home advantage, such as e-sports, so a  comparison of the different models would only be fair if we did not include this property. However, in Section~\ref{sec:Concl} we
formulate the home advantage adaptation given its potential interest.

If pairs of teams do not play equally home and away, then ignoring home advantage could lead to misinterpretation of the estimated ICBT model parameters, e.g., if team $i$ mostly played team $k$ with team $i$ at home, the home advantage would feed into $\theta_{ik}$. We do not believe this is problematic due to the near perfect balance of home to away matches per team, and the maximum home percentage within pairs of teams is 70\% for 2010-11 and only 57\% subsequently.  
%There is no systematic changing of teams from season to season, such as promotion or relegation, which allows for easier comparison of teams.

\subsection{Inference}
\label{sec:baseball_inf}
The baseball data are analysed using the ICBT model, and its results are compared with those of the  Bradley-Terry model and with the existing models of Section \ref{sec:litBT} except for the model of \cite{duan2017generalized} due to the subjective choices required for some parameters.

\begin{figure}[h!]
\begin{minipage}[t]{0.42\linewidth}
%\centering
%intrans_plot_v2_2018
\includegraphics[scale=0.53]{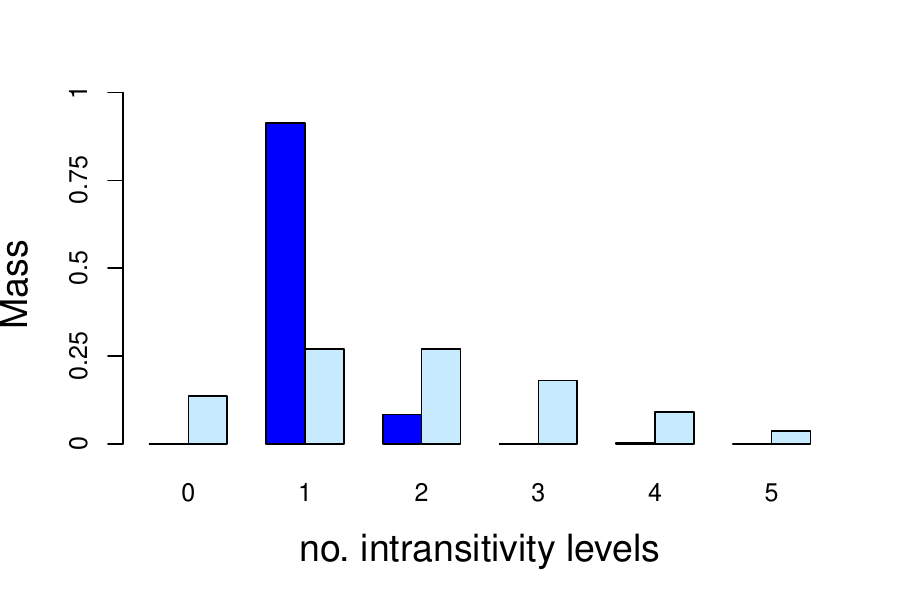}
\end{minipage}
\hfill%
%\fbox{%
\begin{minipage}[t]{0.48\linewidth}
%\centering
%rank_2018
\includegraphics[scale=0.57]{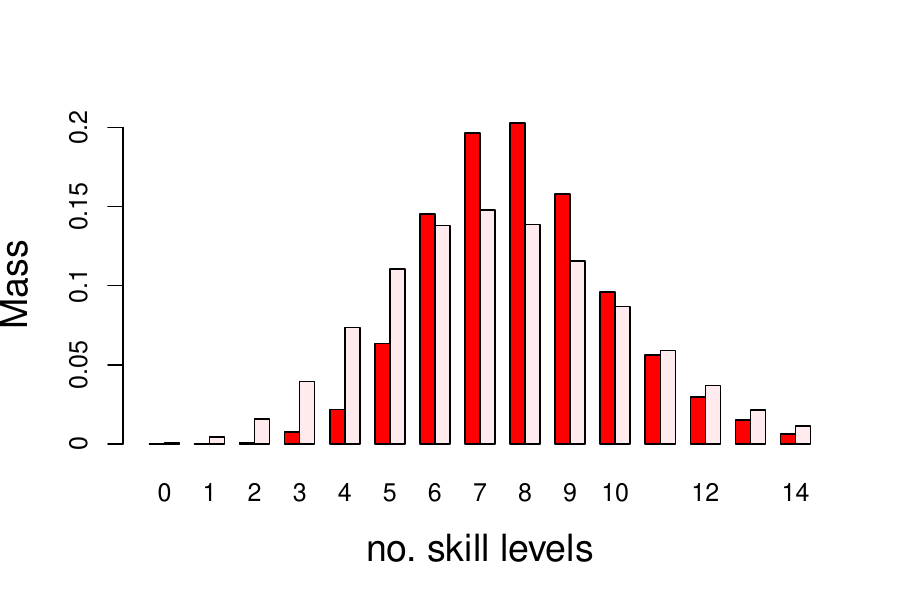}
\end{minipage}
\caption{Posterior distributions of the $K$ intransitivity levels (left) and the $A$ skill levels (right) for the 2018 season: with the associated prior distributions in a lighter colour.}
\label{fig:2018_posterior}
\end{figure}

The ICBT model incorporates uncertainty in the choice of model itself, that is, the number of clusters and therefore how many parameters.  Our prior distributions for number of intransitivity levels $K$ and skill levels $A$ are shown in Figure~\ref{fig:2018_posterior}. We used a Poisson($\lambda_K=2$) prior for $K$, with the 
hyper-parameter to give a 95\% prior chance that $K\in [0,5]$, as it was thought that there would only be a few  different pairwise strategies. Similarly, the prior for $A$ was taken to be Poisson($\lambda_A = 7$) as this hyper-parameter choice gave a 95\% prior chance that $A\in [2,13]$. The justification for our choice of the other hyper-parameters $(\gamma_K, \gamma_A, \alpha,\beta, \nu_A$) and a sensitivity analysis to hyper-parameter choice is reported in the supplementary material.% \citep{spearing2022Supplementarybaseball}.

Now consider the posterior distributions for $K$ and $A$ based on the 2018 season data, also shown in  
Figure~\ref{fig:2018_posterior}. Despite the prior only providing vague information across values of $K\le5$, 
the data clearly favours having a single intransitivity level, meaning three possible clusters for each pair: a positive level, the completely transitive level, and the mirrored negative level. Further, although the prior gave a 14\% probability to the Bradley-Terry model $(K=0)$, the posterior probability for that model is estimated to be zero, showing strong evidence of intransitivity in the dataset. For the distinct skill levels the change from prior to posterior is relatively small, with a mean (and 95\% credible interval) of $7.94\; (4,12)$, with the number of distinct skills levels favouring $A\in [6,9]$.

The posterior estimated values of the teams' skills, and the variance of these values in particular, provides a helpful summary of how competitive the tournament is in a season, with the smaller the variance the more closely contested the tournament. For our model the skills' variance ranged from $0.027$ (2012 season) to $0.32$ (2018 season) and for the standard Bradley-Terry model, $0.031$ (2015 season) to $0.23$ (2018 season) - both models suggesting that 2018 was the least competitive season. %The largest discrepancy between the models was found in the in the 2013 season, where the variance of skill levels according to Bradley-Terry is almost $3$ times that of our model. 
In the 2013 season the variance of skill levels according to Bradley-Terry is almost $3$ times that of our model. However, the 2013 season was found to contain a particularly large amount of intransitivity, indicating that the large range of skills in the Bradley-Terry model could be the result of compensating for an inability to express the intransitivity. This has perhaps resulted in the Bradley-Terry model concluding that the 2013 season was less competitive than it was in reality.

\begin{figure}[h!]
\begin{minipage}[t]{0.42\linewidth}
%\centering
%intrans_plot_v2_2018
\includegraphics[scale=0.53]{plots/baseball/intrans_plot_rev_2018.pdf}
\end{minipage}
\hfill%
%\fbox{%
\begin{minipage}[t]{0.48\linewidth}
%\centering
%rank_2018
\includegraphics[scale=0.57]{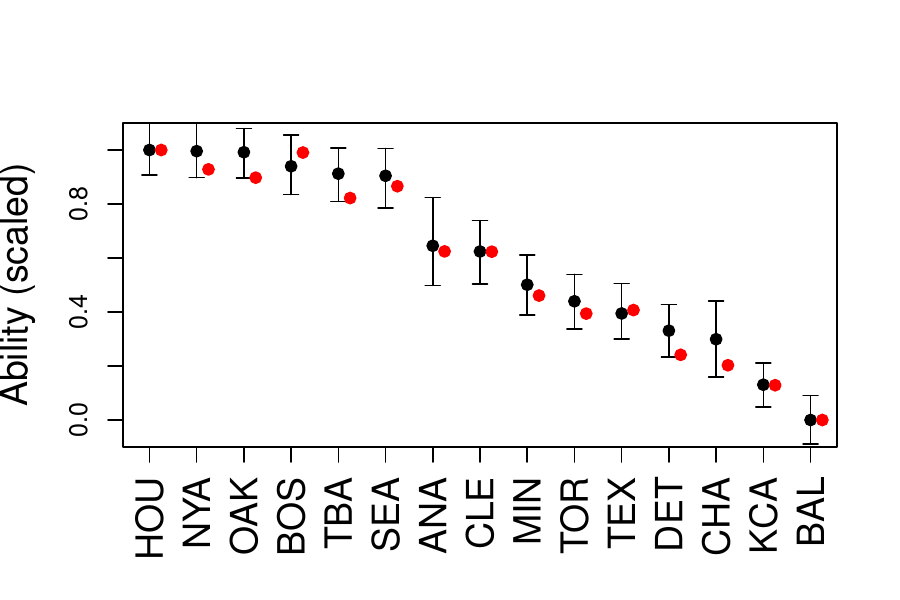}
\end{minipage}
\caption{Analysis of 2018 season: the posterior mean of the intransitivity parameter, $\hat{\theta}^*_{ij}$ across all pairs of teams $i>j \in \mathcal{I}$ (left); ranking according to definition \eqref{eq:ranking_p} (black) and Bradley-Terry model (red) for all teams $i\in\mathcal{I}$ (right).}
\label{fig:2018_ranking}
\end{figure}

Now consider the pairwise interactions between teams. These interactions could be inferred from either the \textit{intransitivity of the posterior mean} 
$\hat{\theta}^*_{ij},\;\forall i \neq j$,
%see definition \eqref{eq:intrans2}, 
or by the \textit{posterior mean of the intransitivity parameter} $\hat{\theta}_{ij},\;\forall i \neq j$. The supplementary material contains a comparison
%\cite{spearing2022Supplementarybaseball} 
 for both and concludes that $\hat{\theta}^*_{ij}$ is more meaningful and interpretable here, so we focus on that. For the 2018 season, Figure~\ref{fig:2018_ranking} (left) shows $\hat{\theta}^*_{ij}$, for each pair of teams $i > j \in \mathcal{I}$: recall that intransitivity has rotational symmetry, i.e., ${\theta_{ij}^* = -\theta_{ji}^*,\;\forall i\neq j}$. The teams are sorted by their rank according to ${p_.}$, given by definition \eqref{eq:ranking_p}, see Figure \ref{fig:2018_ranking} (right). Reading from the teams on the $y$-axis to $x$-axis there is a large positive value of intransitivity from Baltimore (BAL) to Tampa Bay (TBA) of $0.78$ with 95\% credible interval $(0.36, 1.22)$, indicating that Baltimore played better against Tampa Bay than expected, given their overall abilities. This is consistent with the data, with Baltimore winning 11 out of 19 matches between the two teams, despite being ranked lower.

The analysis of these intransitivities between pairs, and that of the skills of each team, can be combined to produce an overall ranking of the teams. As discussed in Section~\ref{sec:model_interp}, with further details in the supplementary material, %\cite{spearing2022Supplementarybaseball}, 
${p}_.$ provides a suitable ranking of the teams. For the 2018 season Figure \ref{fig:2018_ranking} (right) shows the ranks according to ${p}_.$ compared to the Bradley-Terry ranks. Both have been linearly scaled to help with a visual comparison, such that the best and worst teams have abilities 1 and 0 respectively. The two sets of estimated rankings using ${p}_.$  are clearly correlated; however, there is some difference in the ordering of the ranks, indicating that intransitivity may have been masking the true ranks of some teams. For example, consider Tampa Bay, ranked 6th by the Bradley-Terry model. Tampa Bay's good record against Kansas City (KCA) has a much lower weighting than their poor record against Baltimore in the Bradley-Terry model due to the differing frequency of these match-ups, and therefore impacts the overall rank of Tampa Bay. The ICBT model however, recognises that good or bad records against particular teams could be due to the presence of intransitivity, and therefore penalises Tampa Bay less overall, ranking them 5th, thus illustrating our point in 
Section~\ref{sec:intro} that the ICBT model makes adjustments for tournament imbalance.  Similar plots and inferences are drawn from the other seasons (2010-2017) but with different team rankings in each year. 

\subsection{Model Performance}
\label{sec:baseball_performance}
To test the model performance, 70\% of games from each season are randomly selected to be training data, on which the model is fitted, with the remaining 30\% used as test data, on which the log-loss score is calculated. This random selection is appropriate as none of the models compared take time-dependency into consideration, a feature discussed in the supplementary material.
% \cite{spearing2022Supplementarybaseball}. 
 The variation due to this random sampling in the training-test split is accounted for by taking 100 separate random training-test splits for each season. For each replicate of training data the model is fitted separately to each season's data. Relative log-loss is then calculated by subtracting the log-loss of a baseline coin tossing model.

Table \ref{tab:Coin1} shows these negative relative log-loss scores for all years of baseball data, along with 95\% confidence intervals, with this measure evaluated for the ICBT, Bradley-Terry, blade-chest and majority vote models. Since a larger value of negative relative log-loss indicates better model performance, a positive value indicates an improvement on the coin tossing model. So all four models improve on simply using coin tossing, showing that there is information to be exploited for inference and  prediction. The Bradley-Terry, blade chest and majority vote models all have somewhat similar performance to each other across the years, with the most improved fit being in 2018.
In contrast our model is the best performing out of the four models in terms of out-of-sample prediction on all years of data. When assessed as the cumulative improvement over years, relative to coin tossing, the ICBT model is 2.8 times better than the Bradley-Terry model, showing that we have substantially improved predictive performance.
The difference in log-loss scores relative to the Bradley-Terry model is largest for the 2013 season, which in Section~\ref{sec:baseball_inf} has been identified as the season with the largest intransitivity.

\begin{table}[!htbp] \centering 
  \label{tab:Coin1} 
\begin{tabular}{@{\extracolsep{5pt}} ccccc} 
\\[-1.8ex]\hline 
%\hline 
year & ICBT & BT & blade-chest & majority vote \\ 
\hline 
2010 & 44(38, 46) & 17(15, 18) & 17(2, 27) & 20(13, 25) \\ 
2011 & 46(33, 49) & 15(13, 17) & 17(-1, 26) & 20(13, 26) \\ 
2012 & 49(44, 53) & 14(11, 16) & 24(8, 33) & 22(14, 31) \\ 
2013 & 64(36, 69) & 23(21, 25) & 33(13, 42) & 31(22, 39) \\ 
2014 & 39(29, 45) & 9(7, 11) & 10(-12, 21) & 13(6, 19) \\ 
2015 & 34(12, 45) & 5(2, 7) & 9(-14, 17) & 9(1, 16) \\ 
2016 & 42(32, 55) & 10(8, 12) & 18(2, 30) & 18(10, 28) \\ 
2017 & 36(10, 50) & 13(11, 15) & 13(-4, 22) & 16(9, 22) \\ 
2018 & 73(65, 79) & 46(44, 47) & 48(19, 56) & 51(43, 57) \\ 
\hline \\[-1.8ex] 
\end{tabular} 
\caption{Negative relative log-loss $\times 10^3$ (compared to a coin-tossing   model) for each year of baseball data for the ICBT, Bradley-Terry, blade-chest and majority vote models. 95\% confidence intervals, in parentheses, come from random training-test splits of the data.} 
\end{table}

\section{Conclusions and Discussion}
\label{sec:Concl}
We have proposed a new model and inference structure for paired comparison data. We frame this in the context of sport competitions, baseball in particular, with {\it teams} competing against each other, though the potential applications of the model are much broader. Our proposed model, the Intransitive Clustered Bradley-Terry (ICBT) model, extends the standard Bradley Terry model, which is widely considered as the baseline model for such data. The extension allows for intransitivity so that the difference in {\it skill} levels between two objects being compared is not the only factor affecting the probabilities of the outcomes. There are a number of models which already allow for intransitivity, but each of these are quite restricted in the parametric form of intransitivity relative to our semi-parametric approach, which recognises that certain patterns of interaction between pairs of objects can be common over multiple pairs. Our model also allows for objects’ skills to be clustered, a feature that is novel to paired comparisons, with this inducing parsimony and avoiding obtaining distinct rankings for some items when there is no evidence from the data that they are not equally good.  We have shown evidence from American League baseball that our model provides a distinct improvement on existing models. 

The ICBT model has complete flexibility, in the sense that cluster allocation to skill and intransitivity levels is not predetermined. In order that the data identify the appropriate structure of clustering, and for the inference to account for the uncertainty in this choice, the model is fitted via RJMCMC.  

Based on the clusters with the highest posterior probabilities, we anticipate that experts in the particular sport may be able to identify some patterns of clustering that are interpretable, e.g., associated with different styles of play. In such cases, these clustering features could be hard wired into the model as the only options, resulting in more efficient inference.  A referee made the helpful suggestion that if accounting for clustering uncertainty was not an issue then the inference could be simplified by estimating the ICBT model with group lasso penalties to induce clusters. We feel that our model works sufficiently well for the current applications but agree that it presents an exciting springboard for the consideration of various extensions to the model and its inference. We finish by illustrating a few such possible extensions. 

In Section~\ref{sec:baseball} we did not attempt to account for home advantage, which is widely recognised as an important feature in sport, e.g., \citet{cattelan2013dynamic} incorporate it in a Bradley-Terry model, though to the best of our knowledge it has not been accounted for in the existing intransitivity models. The most natural way to achieve this is to change $p_{ik}$ given by expression~\eqref{eq:IBT2} to
a probability $p^{(i)}_{ik}$ of the home team $i$ beating the away team $k$, with
\begin{equation}
\label{eq:IBTH}
p^{(i)}_{ik} = \frac{1}{1 + \exp[-\left(\theta_{ik} +\gamma+r_{i} - r_{k}\right)]},
\mbox{ and }p^{(i)}_{ki}=1-p^{(i)}_{ik}  \;\forall i \neq k \in \mathcal{I},
\end{equation}
where $\gamma \in \mathbb{R}$ determines the effect of playing at home, which here is  common over all pairs of teams. If  $\gamma>0$ ($\gamma<0$) then the probability of a home win is increased (decreased) relative to the other factors of skill and intransitivity. This effect can be extended to vary over teams by replacing $\gamma$ by $\gamma_i$ in expression~\eqref{eq:IBTH}.
To ensure these $\gamma_i$ parameters are all identifiable, we fix $\gamma_1=0$, though no additional constraints are needed if there is a common $\gamma$, but that is all that is required under the conditions of Proposition~\ref{prop:ID} on the other parameters, as we are able to exploit data that distinguishes which team is at home.

This article only considered win-loss scenarios. Extensions of the Bradley-Terry have been proposed for handling draws. Two distinct methods for handling draws are given by \citet{cattelan2013dynamic} and \citet{hankin2020generalization}.
The former use ordinal logistic regression, treating win, loss, draw as outcomes of an ordered multinomial random variable, which can then be analysed via an ordered link model. In contrast, the latter treats the problem as a competition between the two teams and a third theoretical team, such that when the theoretical team wins the outcome of the match corresponds to a draw between the two actual teams. The ICBT model can be adapted similarly, with the use of the clustering strategy extended to pooling teams to account for their similar cautiousness, leading to them drawing more often than would be expected. 

We have assumed that all teams play each other. If this is not the case we cannot improve on the prior inference for the $\theta_{ik}$ parameters for pairs $(i,k)$ that do not play each other. This is not a restriction for Bradley-Terry or the existing intransitivity models, where 
the associated $p_{ik}$ are determined by the observed pairs. This raises issues about identifiability of the ICBT model parameters.  Our approach, through Proposition~\ref{prop:ID}, is no longer sufficient leaving the open problem of which parameters to fix in order to give the most efficient inference.

\subsection*{Acknowledgements}
We thank Dr.\ Matthew Ludkin for giving detailed insight into the reversible jump algorithm. Spearing gratefully acknowledges funding of the EPSRC funded STOR-i Centre for Doctoral Training
(grant number EP/L015692/1), and ATASS Sports. We thank the referees for their helpful comments.
\subsection*{Disclosure Statement}
The authors report there are no competing interests to declare.

\subsection*{Supplementary Materials}
\begin{description}
\item[Additional Analyses:] The supplementary material for this article include extra analyses: a simulation study; and an extended analysis of the baseball application, including a pooled analysis of data from all seasons. (\verb!ICBT_supplementary.pdf!)

\item[Reversible Jump Markov Chain Monte Carlo Sampler:] The supplementary material includes full details of the RJMCMC algorithm, including the initialisation procedure and highlighting the main novelties of the algorithm. (\verb!ICBT_supplementary.pdf!)

\item[R Code:] The analyses can be replicated using the code and data found in the supplementary material. See the README contained in the zip file for more details. (\verb!spearing_ICBTcode.zip!, zip archive)

\item[Appendix:] The supplementary material contains the Appendix, which provides the proof of Proposition \ref{prop:ID}. (\verb!ICBTappendix.pdf!)
\end{description}

\linespread{1}
%\newpage
\bibliographystyle{apalike}
\bibliography{ICBT_model}

\begin{thebibliography}{}

\bibitem[Bradley and Terry, 1952]{bradley1952rank}
Bradley, R.~A. and Terry, M.~E. (1952).
\newblock Rank analysis of incomplete block designs: I. the method of paired
  comparisons.
\newblock {\em Biometrika}, 39(3/4):324--345.

\bibitem[Cattelan et~al., 2013]{cattelan2013dynamic}
Cattelan, M., Varin, C., and Firth, D. (2013).
\newblock Dynamic {B}radley--{T}erry modelling of sports tournaments.
\newblock {\em Journal of the Royal Statistical Society: Series C (Applied
  Statistics)}, 62(1):135--150.

\bibitem[Causeur and Husson, 2005]{causeur20052}
Causeur, D. and Husson, F. (2005).
\newblock A 2-dimensional extension of the {B}radley--{T}erry model for paired
  comparisons.
\newblock {\em Journal of Statistical Planning and Inference}, 135(2):245--259.

\bibitem[Chen and Joachims, 2016]{chen2016modeling}
Chen, S. and Joachims, T. (2016).
\newblock Modeling intransitivity in matchup and comparison data.
\newblock In {\em Proceedings of the Ninth ACM International Conference on Web
  Search and Data Mining}, pages 227--236. ACM.

\bibitem[Dixon and Coles, 1997]{dixon1997modelling}
Dixon, M.~J. and Coles, S.~G. (1997).
\newblock Modelling association football scores and inefficiencies in the
  football betting market.
\newblock {\em Journal of the Royal Statistical Society: Series C (Applied
  Statistics)}, 46(2):265--280.

\bibitem[Duan et~al., 2017]{duan2017generalized}
Duan, J., Li, J., Baba, Y., and Kashima, H. (2017).
\newblock A generalized model for multidimensional intransitivity.
\newblock In {\em Pacific-Asia Conference on Knowledge Discovery and Data
  Mining}, pages 840--852. Springer.

\bibitem[Green, 1995]{green1995reversible}
Green, P.~J. (1995).
\newblock Reversible jump {M}arkov chain {M}onte {C}arlo computation and
  {B}ayesian model determination.
\newblock {\em Biometrika}, 82(4):711--732.

\bibitem[Green and Richardson, 2001]{green2001modelling}
Green, P.~J. and Richardson, S. (2001).
\newblock Modelling heterogeneity with and without the {D}irichlet process.
\newblock {\em Scandinavian Journal of Statistics}, 28(2):355--375.

\bibitem[Hankin, 2020]{hankin2020generalization}
Hankin, R.~K. (2020).
\newblock A generalization of the {B}radley--{T}erry model for draws in chess
  with an application to collusion.
\newblock {\em Journal of Economic Behavior \& Organization}, 180:325--333.

\bibitem[Ludkin, 2020]{ludkin2020inference}
Ludkin, M. (2020).
\newblock Inference for a generalised stochastic block model with unknown
  number of blocks and non-conjugate edge models.
\newblock {\em Computational Statistics \& Data Analysis}, 152:107051.

\bibitem[Makhijani and Ugander, 2019]{makhijani2019parametric}
Makhijani, R. and Ugander, J. (2019).
\newblock Parametric models for intransitivity in pairwise rankings.
\newblock In {\em The World Wide Web Conference}, pages 3056--3062. ACM.

\bibitem[Makowski and Piotrowski, 2006]{makowski2006quantum}
Makowski, M. and Piotrowski, E.~W. (2006).
\newblock Quantum cat's dilemma: an example of intransitivity in a quantum
  game.
\newblock {\em Physics Letters A}, 355(4-5):250--254.

\bibitem[Masarotto and Varin, 2012]{masarotto2012ranking}
Masarotto, G. and Varin, C. (2012).
\newblock The ranking lasso and its application to sport tournaments.
\newblock {\em The Annals of Applied Statistics}, 6(4):1949--1970.

\bibitem[Smead, 2019]{smead2019sports}
Smead, R. (2019).
\newblock Sports tournaments and social choice theory.
\newblock {\em Philosophies}, 4(2):28.

\end{thebibliography}
\end{document}